\documentclass[12pt,preprint]{aastex}
\usepackage{amssymb}
\usepackage{epsfig}

\begin{document}


\title{Cold Molecular Gas Along the Cooling X-ray Filament in Abell~1795}


\author{Michael McDonald$^{1,5}$, Lisa H. Wei$^{2,3,7}$, Sylvain Veilleux$^{4,5,8}$ }
\altaffiltext{1}{Kavli Institute for Astrophysics and Space Research, MIT, Cambridge, MA 02139, USA}
\altaffiltext{2}{Harvard-Smithsonian Center for Astrophysics, 60 Garden Street, Cambridge, MA 02138, USA}
\altaffiltext{3}{Atmospheric and Environmental Research, 131 Hartwell Avenue Lexington, MA  02421}
\altaffiltext{4}{Department of Astronomy, University of Maryland, College Park, MD 20742, USA}
\altaffiltext{5}{Astroparticle Physics Laboratory, NASA Goddard Space Flight Center, Code 661, Greenbelt, MD 20771 USA}
\altaffiltext{6}{Email: mcdonald@space.mit.edu}
\altaffiltext{7}{Email: lisa.wei@cfa.harvard.edu}
\altaffiltext{8}{Email: veilleux@astro.umd.edu}


\begin{abstract}

We present the results of interferometric observations of the
cool core of Abell 1795 at CO(1-0) using the Combined Array for Research in
Millimeter-Wave Astronomy. In agreement with previous work, we detect
a significant amount of cold molecular gas (3.9 $\pm$ 0.4
$\times$10$^9$ M$_{\odot}$) in the central $\sim$10 kpc. We report the
discovery of a substantial clump of cold molecular gas at
clustercentric radius of 30 kpc (2.9 $\pm$ 0.4 $\times$10$^9$
M$_{\odot}$), coincident in both position and velocity with the warm,
ionized filaments. 
We also place an upper
limit on the H$_2$ mass at the outer edge of the star-forming
filament, corresponding to a distance of 60 kpc ($<$0.9 $\times$10$^9$
M$_{\odot}$). We measure a strong gradient in the H$\alpha$/H$_2$ ratio as a function of radius, suggesting different ionization mechanisms in the nucleus and filaments of Abell1795. The total mass of cold molecular gas ($\sim$7$\times$10$^{9}$
M$_{\odot}$) is roughly 30\% of the classical cooling estimate at
the same position, assuming a cooling time of 10$^9$ yr.  Combining
the cold molecular gas mass with the UV-derived star formation rate
and the warm, ionized gas mass, the spectroscopically-derived X-ray
cooling rate is fully accounted for and in good agreement with the
cooling byproducts over timescales of $\sim$10$^9$ yr.  The overall
agreement between the cooling rate of the hot intracluster medium and
the mass of the cool gas reservoir suggests that, at least in this
system, the cooling flow problem stems from a lack of observable
cooling in the more diffuse regions at large radii.
\end{abstract}




\section{Introduction}

The search for cooling gas in the cores of galaxy clusters has a rich history. Since it was first discovered that the intracluster medium (ICM) in some galaxy clusters is cooling on timescales shorter than the age of the Universe \citep[][]{lea73, cowie77, fabian77, mathews78}, considerable time and effort has been spent searching for the byproducts of this cooling. Surveys to detect line emission from cooling gas at $\sim$10$^5$--10$^6$~K \citep[OVI; e.g.,][]{bregman01, oegerle01a, bregman06}, $\sim$10$^4$~K, \citep[H$\alpha$; e.g.,][]{hu85, heckman89, crawford99, edwards07, mcdonald10}, $\sim$10$^3$~K \citep[warm H$_2$; e.g.,][]{jaffe97, donahue00,edge02,hatch05,jaffe05},  $\sim$10$^1$--10$^2$~K \citep[{[}\ion{C}{2}{]}; e.g.,][]{edge10, mittal11}, and $\sim$10$^1$~K \citep[CO; e.g.,][]{edge01,edge03,salome03,salome08} have continued to return the same result: there is considerably less cool gas in cluster cores than predicted by radiative cooling of the hot ICM.  It is now assumed that one or a combination of energetic processes balance radiative cooling, allowing only a small fraction of the cooling ICM to reach temperatures below $\sim$10$^6$~K \citep[e.g.,][]{peterson03, peterson06}. Leading candidates for this source of heating are cosmic rays \citep[e.g.,][]{ferland08,mathews09,fujita11} and feedback from the central active galactic nucleus \citep[AGN; e.g.,][]{mcnamara07, guo08, ma11,randall11}. 

Assuming some fraction of the ICM cools unimpeded, it will eventually wind up as cold molecular gas. While much work has gone into quantifying the total mass of the cold gas reservoir in cluster cores \citep[e.g.,][]{edge01,edge03,salome03}, it is considerably more challenging to study the detailed morphology of this gas. Only the Perseus cluster, due to its proximity, has CO detected with the same spatial extent as the soft X-ray- and optically-emitting gas \citep{hatch05, salome11}. Mapping the distribution of the cold gas is crucial to understanding the complicated relation between cooling and feedback processes, and in estimating the amount of fuel for star formation and black hole growth.

In this Letter, we present recent high spatial resolution interferometric observations of CO(1-0) in Abell 1795 (hereafter A1795) from the Combined Array for Research in Millimeter-wave Astronomy (CARMA). This well-studied cluster has a pair of 60 kpc long cooling filaments detected in X-ray \citep{crawford05}, H$\alpha$ \citep[e.g.,][]{cowie83,mcdonald10}, and far-UV \citep{mcdonald09}. Our observations cover the full extent of the cooling filaments, providing a first estimate of the efficiency of gas cooling from the hot phase to the cold phase removed from the influence of the AGN. In \S2 we present these data, briefly describing their acquisition and reduction. In \S3 we present the results of these observations, identifying regions with CO(1-0) line emission and determining the total mass in molecular gas. These results are discussed in the context of the cooling flow model, and compared to previous observations of A1795 and the Perseus cluster in \S4. Finally, in \S5 we conclude with a summary of these results and their implications for future work.

Throughout this study, we assume a luminosity distance to A1795 of 263 Mpc.


\section{CARMA Data}

The CARMA CO(1--0) observations were taken in Nov 2010 and April 2011
in the C array, with a single pointing at the phase center
($\alpha$,$\delta$)$_{\rm J2000}$ = (207.3243$^{\circ}$,+26.5334$^{\circ}$).
At $\sim$108.4\,GHz
(redshifted frequency of CO(1-0) for A1795), the half-power beamwidth
of the 10-meter antennas is $\sim$60$^{\prime\prime}$. T$_{\rm sys}$
ranged from 100--300\,K, with a total on-source time of
15.7\,hours. The correlator was configured to cover a 600\,MHz
bandwidth with 5 overlapping windows (native spectral resolution =
0.4\,MHz, $\sim$1\,km\,s$^{-1}$). Three wideband windows covering a
total bandwidth of 1.5\,GHz were used for phase and amplitude
calibration.

Data were flagged for bad channels, antennas, weather, and pointing
using MIRIAD \citep{sault95}. Bandpass calibration was done on a
bright quasar, 3C~273. We performed atmospheric phase calibrations
with the quasar 1310$+$323 every 20 minutes. Absolute flux calibration
was done using MWC~349A (1.2 Jy), which is unresolved in the C
array. The channels were smoothed and averaged to achieve a final
resolution of 25\,km\,s$^{-1}$.  We combine all the data in MIRIAD,
using natural weighting and an additional weighting by noise as
estimated by the system temperature. The data cubes were cleaned to a
cutoff of 2\,$\sigma$ in the residual image. The resulting synthesized
beam is 1\farcs66$\times$1\farcs5, and the resulting image has RMS
$\sim$ 2.5 mJy/beam per channel.

\section{Results}

\begin{figure*}[p]
\centering
\includegraphics[width=0.95\textwidth]{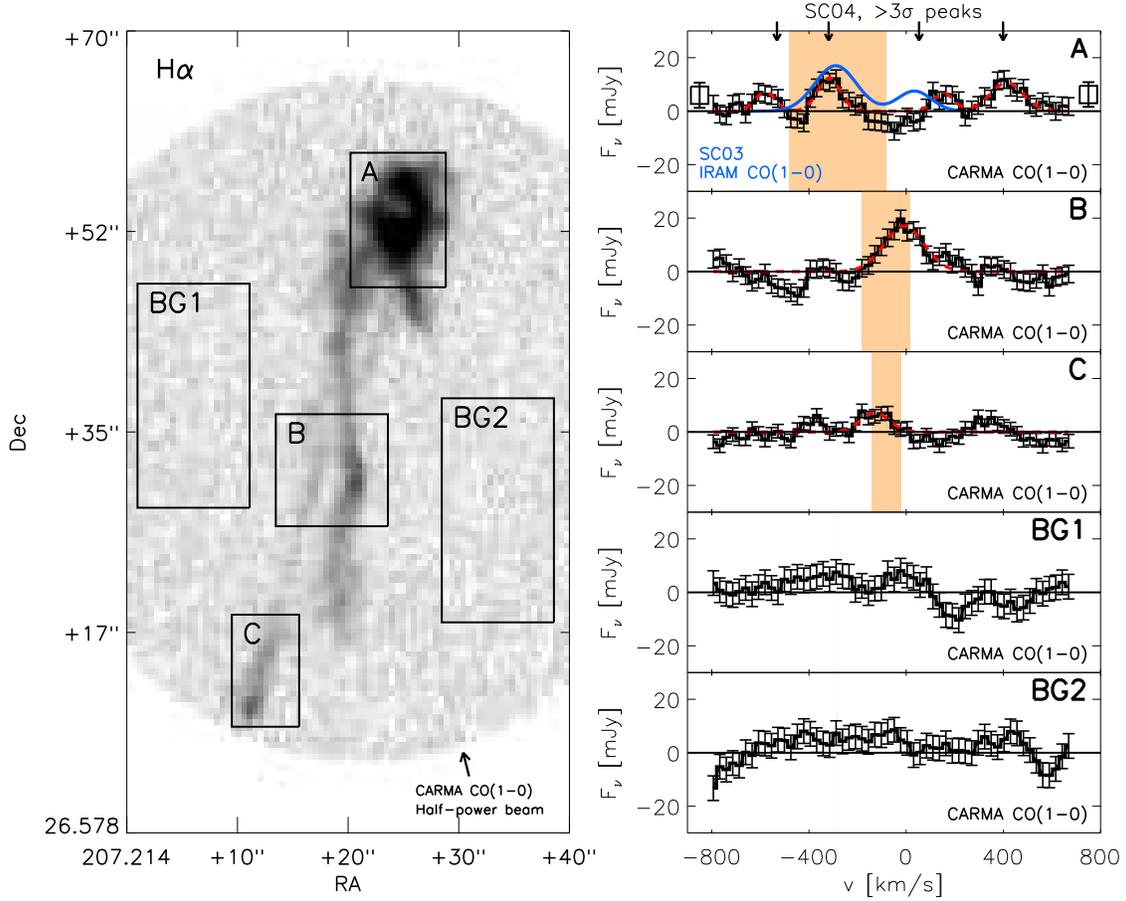}
\caption{Integrated CO(1-0) spectra extracted along the cooling filaments in A1795. The box-size/location, overlaid on an H$\alpha$ image, represents the region over which the CO(1-0) spectra are integrated. Spectra have 25 km s$^{-1}$ bins and are Hanning-smoothed over the adjacent 4 pixels. Flux calibration assumes a conversion factor of K/Jy=37.1. Velocities are relative to the BCG ($z=0.06325$). Panels to the upper left and lower right (BG1, BG2) provide an estimate of the RMS fluctuations in the CO(1-0) spectra. This figure shows four peaks in the central region (A), consistent in velocity with previous work by SC04 (black arrows). We detect an additional strong peak at a cluster-centric distance of $\sim$30~kpc (B) and a velocity of $-$9 km s$^{-1}$.
CO(1-0) velocities in panels A and B agree with the radial velocity of the warm, ionized gas \citep[shaded tan regions;][]{mcdonald12a}, further strengthening the detection significance. At the tip of the filament ($\sim$60 kpc; box C), we place an upper limit on the CO(1-0) flux at the same velocity as the warm gas ($-$129 km s$^{-1}$). Gaussian fits to the spectra are shown as dotted red lines and the flux of the continuum source (measured from wideband data) is shown as two open boxes on the extreme left and right in panel A.}
\label{grid}
\end{figure*}

We expect any detections of molecular gas along the filaments to be only slightly above the noise level based on prior single-dish \citep[][hereafter SC03]{salome03} and interferometric \citep[][hereafter SC04]{salome04} measurements. To increase the signal-to-noise, spectra are extracted in regions centered on the brightest cluster galaxy (hereafter BCG) and along the H$\alpha$-emitting filaments. Figure \ref{grid} shows the CO(1-0) spectra in these three regions, along with adjacent noise estimates. We detect CO(1-0) emission coincident with the BCG center, consistent with earlier work by SC03 and SC04. The CO(1-0) emission in the central region shows four peaks, with velocities, velocity widths, and intensities summarized in Table \ref{table}. The velocities of all four lines are consistent with the four $>$3$\sigma$ detections reported in SC04. The brightest emission line in panel A is consistent with the range of radial velocities for the nuclear warm H$\alpha$-emitting gas \citep{mcdonald12a}. At a distance of $\sim$30~kpc from the BCG center, we detect a second concentration of CO, coincident with the H$\alpha$ filaments in both position and velocity. 
Despite the fact that the H$\alpha$ emission is an order-of-magnitude fainter at a radius of $\sim$30~kpc, we find roughly equal amounts of CO at 30~kpc as in the cluster center.  Finally, we place an upper limit on the CO(1-0) flux at the southern tip of the cooling filament, at a distance of $\sim$60~kpc from the cluster center. The properties of all CO(1-0) line detections are summarized in Table \ref{table}.
We detect a point source, consistent in position with the 1.4GHz source, with a flux of 6.7mJy, consistent with the 7mJy flux reported by SC04. This source has been masked in the spectral extractions shown in Figure \ref{grid}.  We expect side lobes from this masked source to contribute $<$0.7 mJy to the continuum level.


\begin{table}[tb]
\centering
\begin{tabular}{c c c c c c}
\hline\hline
Region & Line Position & Line width & Peak & S$_{CO}$ & M$_{H_2}$\\
 & [km s$^{-1}$] & [km s$^{-1}$] & [mJy] & [Jy km s$^{-1}$] & [10$^8$ M$_{\odot}$]\\
\hline
A & $-$580 $\pm$ 15 & $<$100 & 7.4 $\pm$ 3.7 & 0.75 $\pm$ 0.25 & 6.1 $\pm$ 2.0\\
A & $-$318 $\pm$ 10 & $<$110 & 12.7 $\pm$ 3.5 & 1.48 $\pm$ 0.27 & 12.1 $\pm$ 2.2\\
A & $+$155 $\pm$ 17 & $<$100 & 6.9 $\pm$ 3.7 & 0.71 $\pm$ 0.25 & 5.8 $\pm$ 2.0\\
A & $+$413 $\pm$ 14 & $<$160 & 10.8 $\pm$ 2.9 & 1.80 $\pm$ 0.32 & 14.7 $\pm$ 2.6\\
\\
B & $-$9 $\pm$ 9 & $<$190 & 17.2 $\pm$ 2.6 & 3.53 $\pm$ 0.35 & 28.8 $\pm$ 2.9\\
\\
C & $-$129 $\pm$ 14 & $<$134 & $\leq$7.4 & $\leq$1.06& $\leq$8.6 \\
\hline
\end{tabular}
\caption{Properties of CO(1-0) emission lines in the nucleus (A) and filaments (B, C) of A1795. Velocities are relative to the BCG ($z=0.06325$). Line widths are shown as upper limits, as the spectra have been binned and smoothed to increase signal-to-noise. Calculations of M$_{H_2}$ assume Galactic X$_{CO}$. }
\label{table}
\end{table}

The total emission in the central region implies a molecular gas mass of M$_{H_2}$ =  3.9 $\pm$ 0.4 $\times$10$^9$ M$_{\odot}$, assuming a Galactic value of X$_{CO}$ ($3\times10^{20}$ cm$^{-2}$ (K  km s$^{-1}$)$^{-1}$), following \cite{edge01}. The peak of the emission lines at $-$318 and +155 km s$^{-1}$ are slightly weaker than those measured from single-dish observations presented in SC03. This is most likely due to line emission which is spatially coincident with the continuum source (see Salome \& Combes 2004), which has been masked.
At radii of 30~kpc and 60~kpc we estimate M$_{H_2}$ =  3.8~$\pm$~0.5~$\times$10$^9$ M$_{\odot}$ and M$_{H_2}$ $\leq$ 0.9 $\times$10$^9$ M$_{\odot}$, respectively.

\begin{figure*}[p]
\centering
\begin{tabular}{cc}

\includegraphics[width=0.35\textwidth]{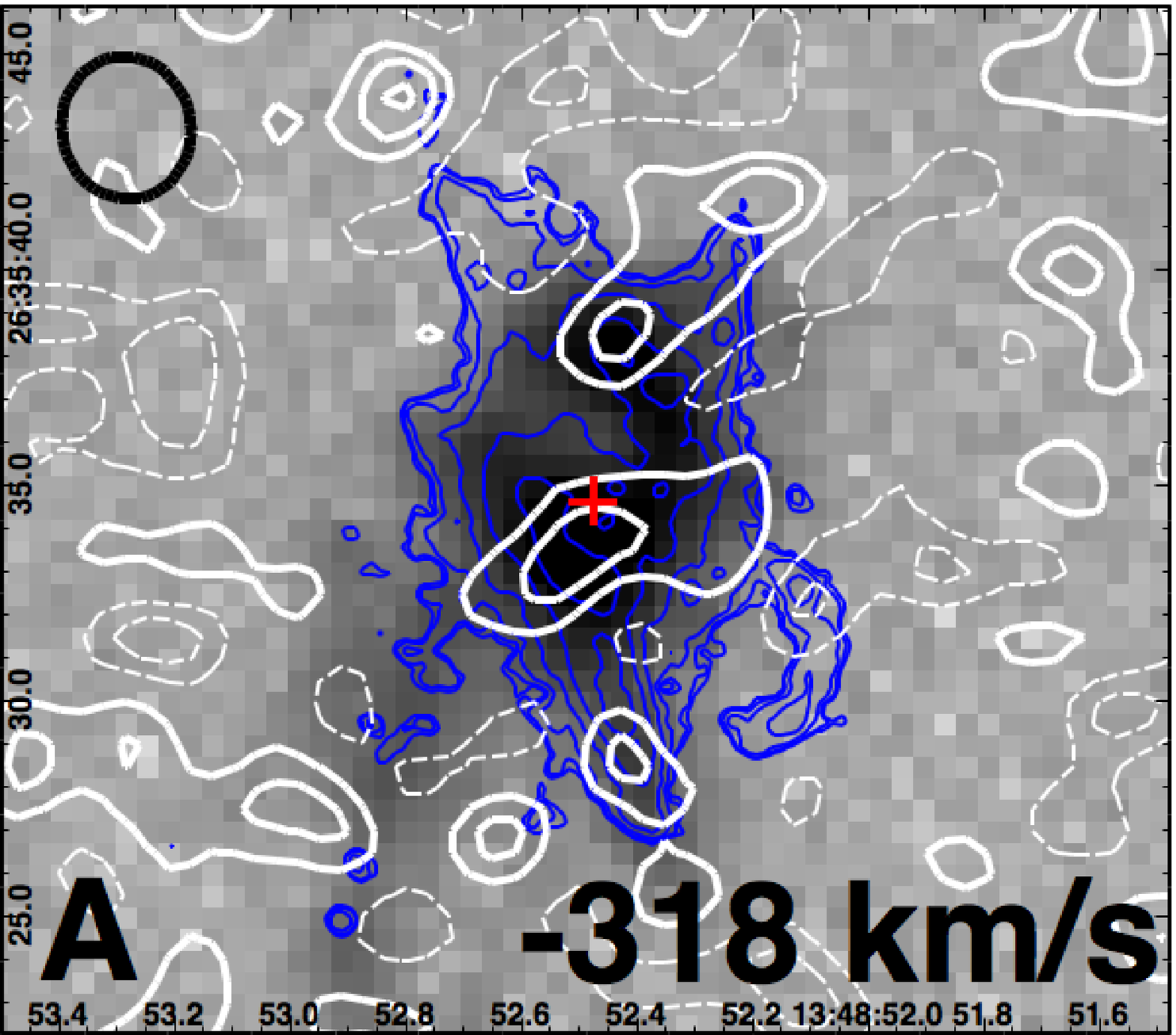} &
~~~~~~~\includegraphics[width=0.35\textwidth]{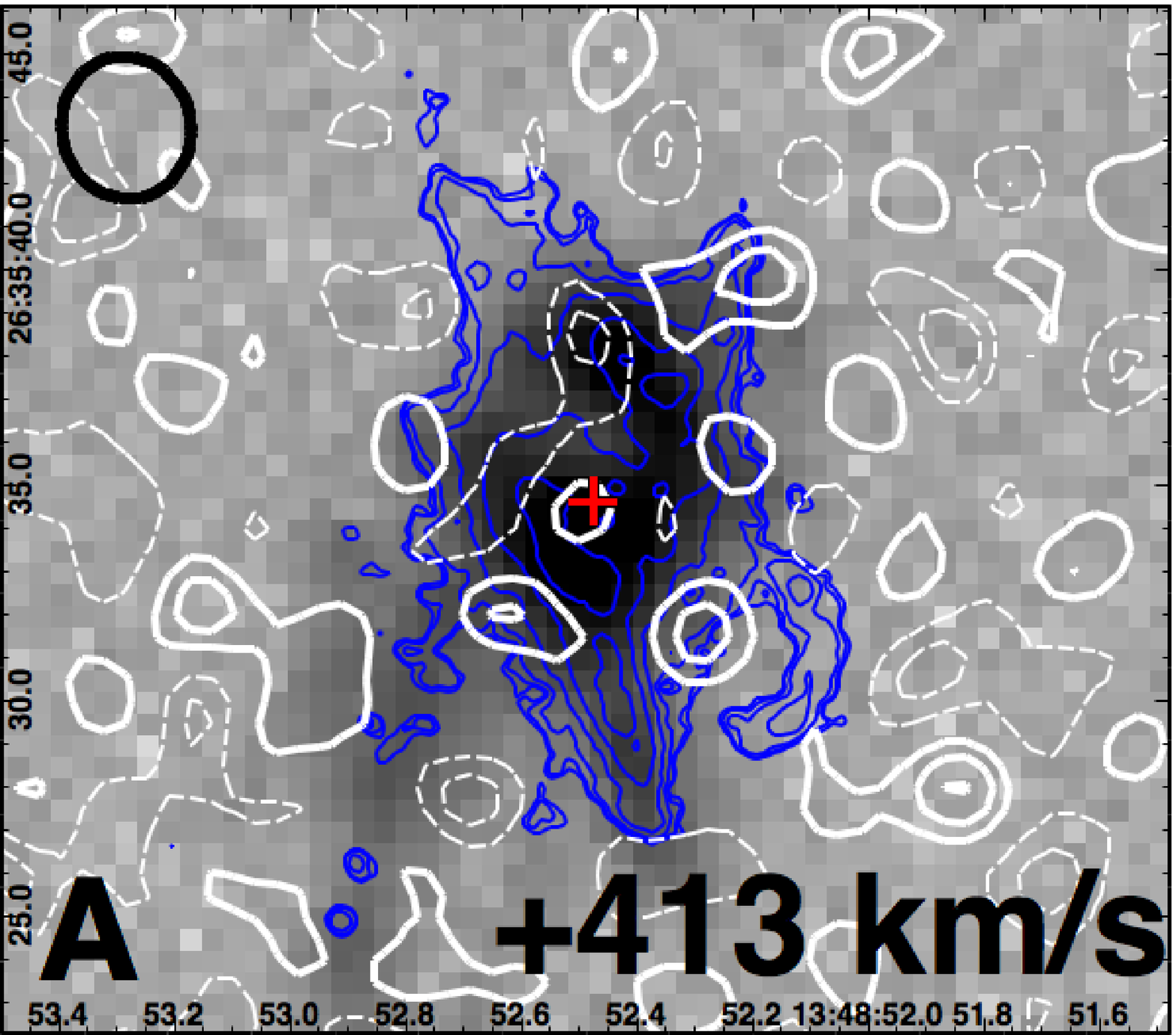} \\
\includegraphics[width=0.35\textwidth]{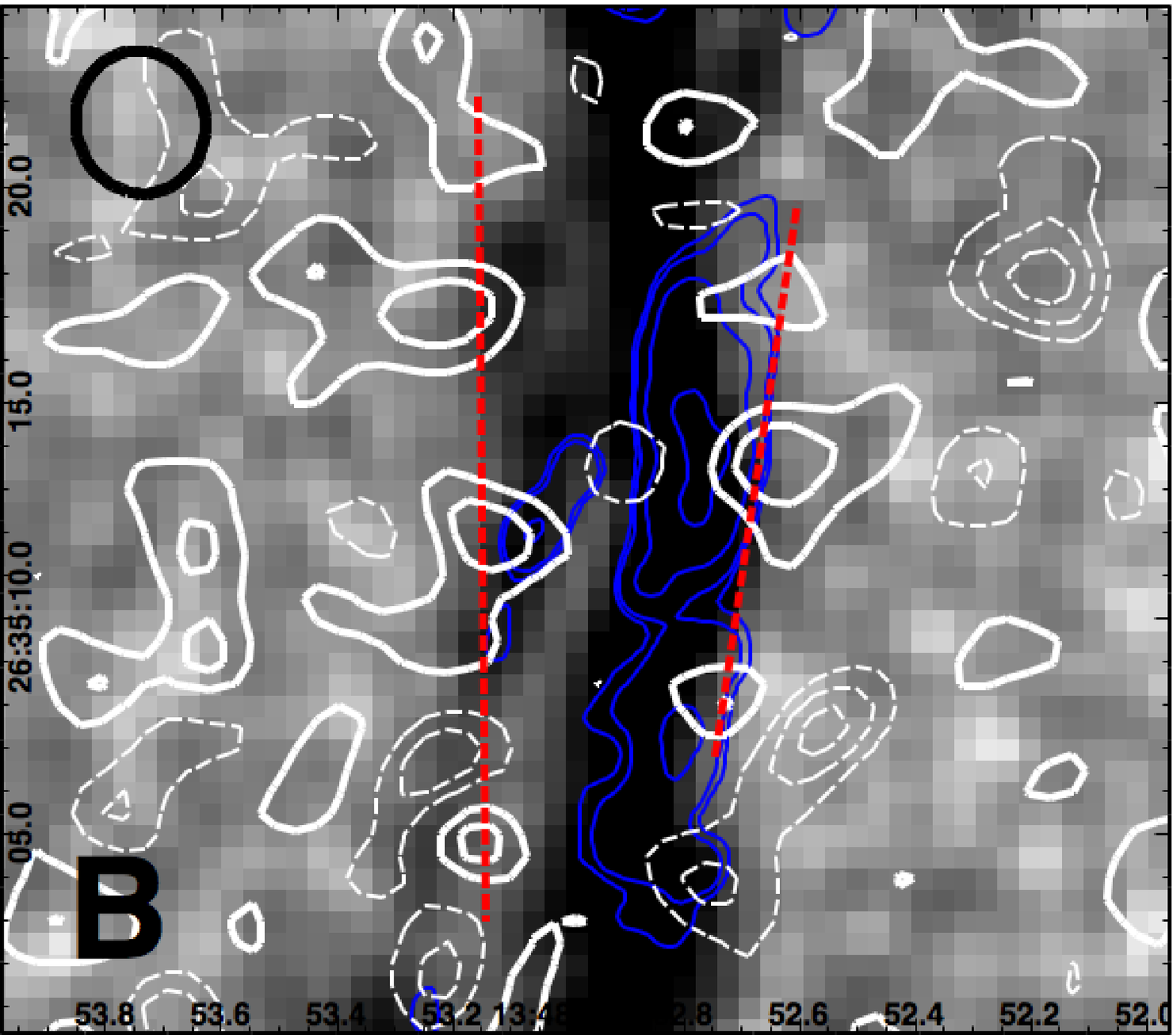} &
\includegraphics[width=0.43\textwidth,trim=0cm 5cm 0cm 0cm]{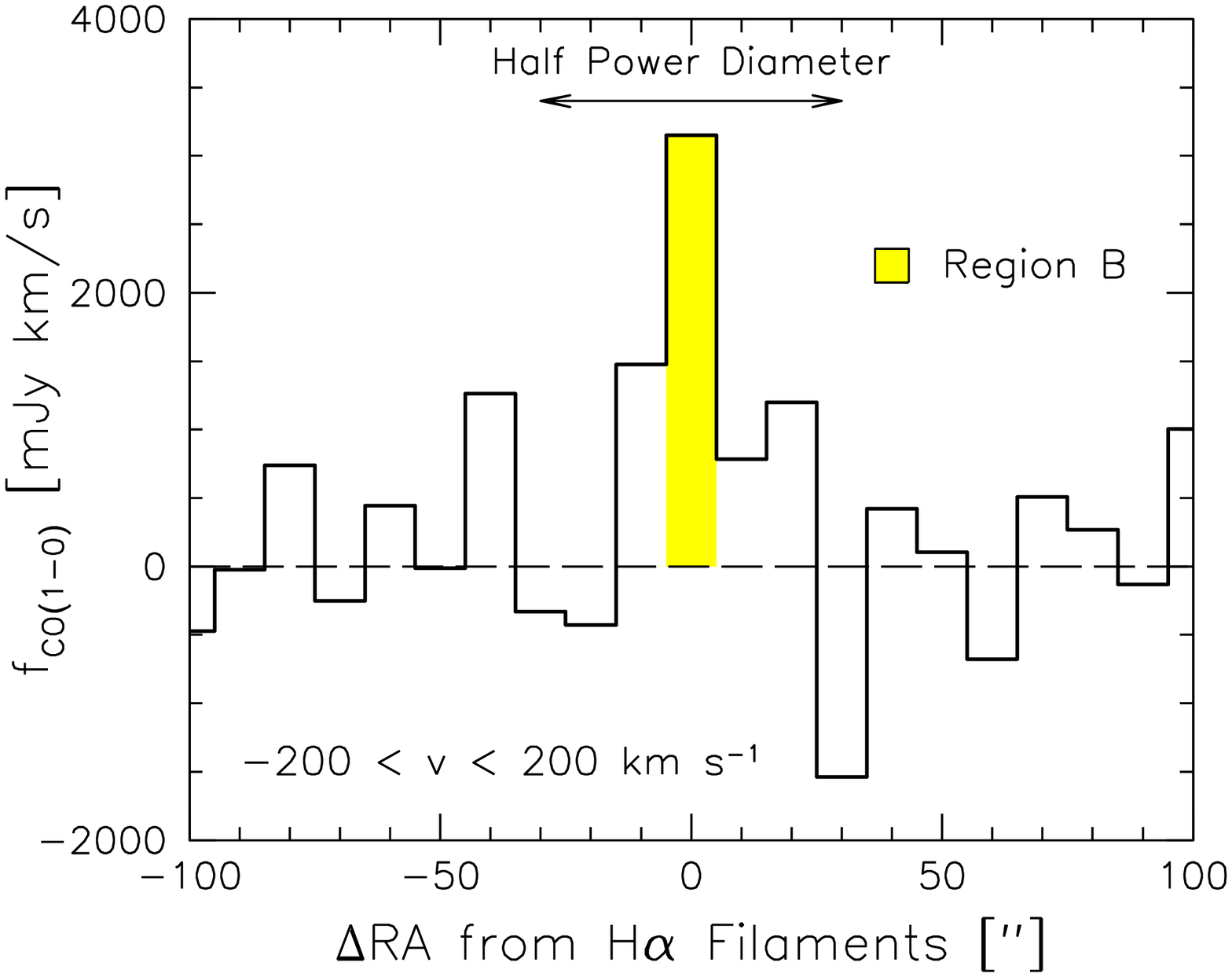} \\

\end{tabular}
\caption{Contour plots showing the velocity-integrated CO(1-0) emission (white) from 3 different emission lines (Fig \ref{grid}). Images are 27$^{\prime\prime}$$\times$24$^{\prime\prime}$. White solid (dashed) contours represent 1, 2, 3, and 4$\sigma$ above (below) zero, where $\sigma$ $\sim$ 0.1 Jy beam$^{-1}$ km s$^{-1}$ in the smoothed images. Black circles in the upper left represent the CO(1-0) beam (1.66$^{\prime\prime}$~$\times$~1.5$^{\prime\prime}$). In general, the morphology of the cold molecular gas matches that of the warm gas (H$\alpha$, greyscale) and young stars (blue contours) at all radii. 
Red dotted lines (panel B) join the emission peaks of the CO(1-0) filaments. 
In the lower right panel, we show a horizontal cut through the full image ($\sim$20$\times$ wider than panel B) with both height and center chosen to match region B from Fig. 1 and a velocity of 0 km/s. This plot shows a significant overdensity of CO(1-0) emission with roughly the same position and velocity as the H$\alpha$ filaments.
}
\label{spatial}
\end{figure*}

Figure \ref{spatial} shows the position of the CO emission compared to the H$\alpha$ and far-UV maps. These maps were created by summing channels over the width of the emission line, centered on the three strongest CO(1-0) peaks separately (Table 1). The strongest emission line in panel A of Figure \ref{grid} originates primarily from gas at the very center of the BCG, coincident with the peak in the X-ray, FUV, optical, and H$\alpha$ emission. This morphology matches SC04, with peaks to the northeast and southwest of the nucleus. The emission line at +400 km s$^{-1}$ appears to originate from slightly larger radii (consistent in position with UV peaks), although the signal-to-noise is insufficient to firmly establish the morphology of this gas. 
The cold gas appears to trace the morphology of the warm gas in the central region, with several 1--2$\sigma$ peaks coincident with the extended H$\alpha$ and far-UV emission. However, while the \emph{integrated} CO(1-0) emission is detected with high ($>$3$\sigma$) significance, Figure \ref{spatial} demonstrates that significantly deeper data is required to carefully map out the morphology of the cold gas.

The cold molecular gas has a similar morphology to the star-forming filaments at a radius of $\sim$30~kpc. While the significance of the individual CO(1-0) peaks is low ($\sim$2$\sigma$), the match in velocity (Fig \ref{grid}) and position (Fig \ref{spatial}) between the H$\alpha$ and CO  suggests that these detections are real. The pair of thin filaments separated by $\sim$4$^{\prime\prime}$ (panel B of Figure \ref{grid}) are each populated by individual chains of CO clumps, with slight ($\sim$1--2$^{\prime\prime}$) offsets between the CO(1-0) and H$\alpha$. This small offset may be due to time delays between the collapse of cold gas clouds and the onset of star formation, or higher optical depths associated with the cold gas, obscuring any coincident optical/UV emission. The majority of the CO(1-0) emission appears to originate in the eastern filament, despite the fact that the western filament is brighter at both H$\alpha$ and FUV intensities, suggesting that, along with being kinematically-distinct \citep{mcdonald12a}, these filaments are at different stages in the cooling process. The fact that the CO(1-0) intensity peaks are roughly consistent in position with the brightest peaks in the H$\alpha$ and FUV maps suggests that CO emission may be detectable along the full extent of both star forming filaments, with a sufficiently deep exposure. In the following section we speculate on what these streams of cold gas imply in the context of the cooling flow model.


\section{Discussion}
\begin{figure*}[t]
\centering
\begin{tabular}{cc}
\includegraphics[width=0.45\textwidth]{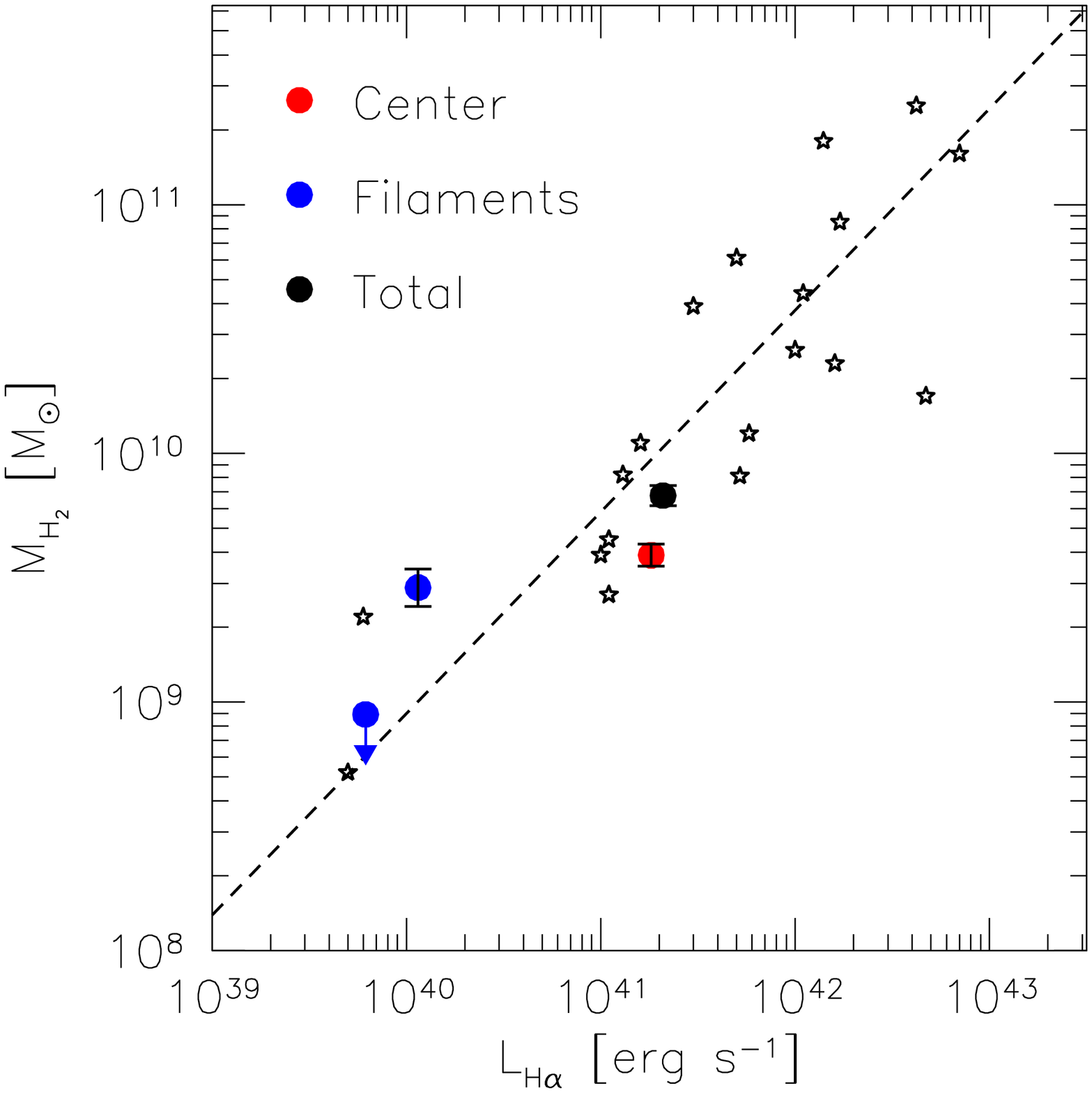} &
\includegraphics[width=0.45\textwidth]{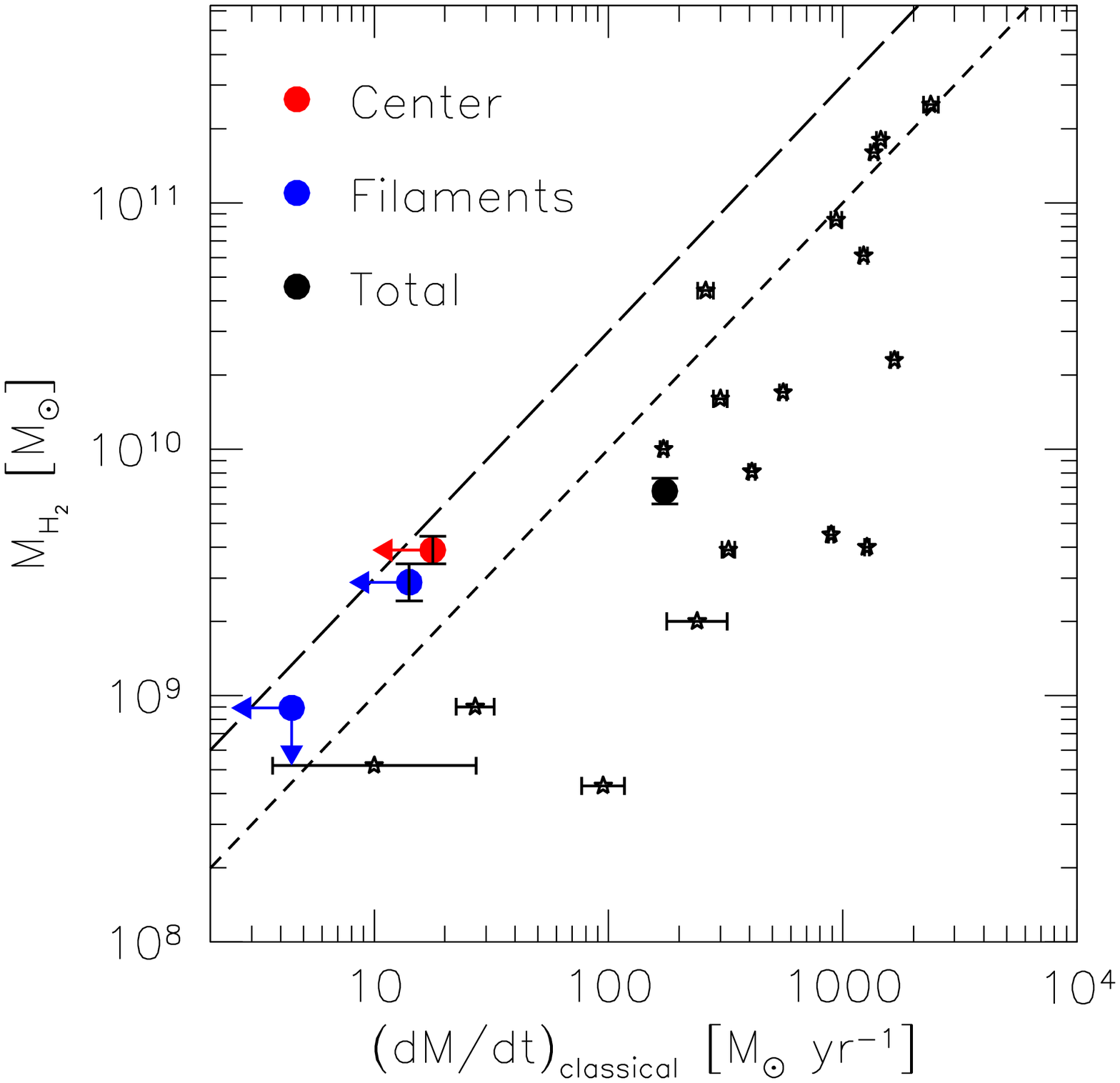} \\
\end{tabular}
\caption{Left: Molecular gas mass (M$_{H_2}$) versus L$_{H\alpha}$ from \cite{edge01} (open stars). Blue and red points represent L$_{H\alpha}$ measured in regions from Figure \ref{grid}. The short-dashed line shows the bisector fit to the data. The filaments appear to be H$_2$-dominated, while the nucleus is H$\alpha$ dominated, suggesting differences in the heating/cooling balance in these regions. Right:  Similar to left panel, but showing the classical cooling rate, (dM/dt)$_{classical}$. Filled circles represent the \emph{local} X-ray cooling rate based on the X-ray luminosity and temperature in the regions from Figure \ref{grid}. These are upper limits since they are uncorrected for projection, which overestimate the cooling rate (due to the increased luminosity from line-of-sight gas). The short- and long-dashed lines represents 10\% and 30\% of the cooling rate over 10$^9$yr.
}
\label{edge}
\end{figure*}

The observation of cold gas at such large distances ($\sim$30 kpc) from the cluster center suggests that the ICM may be cooling efficiently at large radii. To put our estimates of the H$_2$ mass into context, we reproduce two plots from \cite{edge01} in Figure \ref{edge}. The first shows the correlation of the central H$_2$ gas mass with the H$\alpha$ luminosity for a sample of 20 clusters with detections at both CO(1-0) and H$\alpha$. These two quantities are correlated, with A1795 falling along the bisector fit. Considered separately, the filaments appear to be H$_2$ dominant, while the nucleus is H$\alpha$ dominant. This is consistent with the picture presented in \cite{mcdonald12a}, that the optical line emission in the nuclei of cool core BCGs is due to a mix of shocks and star formation, while the filaments typically have a weaker contribution from shocks. Alternatively, the order-of-magnitude difference in L$_{H\alpha}$ between the nucleus and filaments, at a nearly fixed H$_2$ gas mass, may be due to different aged starbursts -- L$_{H\alpha}$ will decrease by a factor of $\sim$100 during the first 10~Myr of a typical starburst.

\begin{figure*}[p]
\centering
\includegraphics[width=\textwidth, trim=1cm 7.cm 0.7cm 0cm]{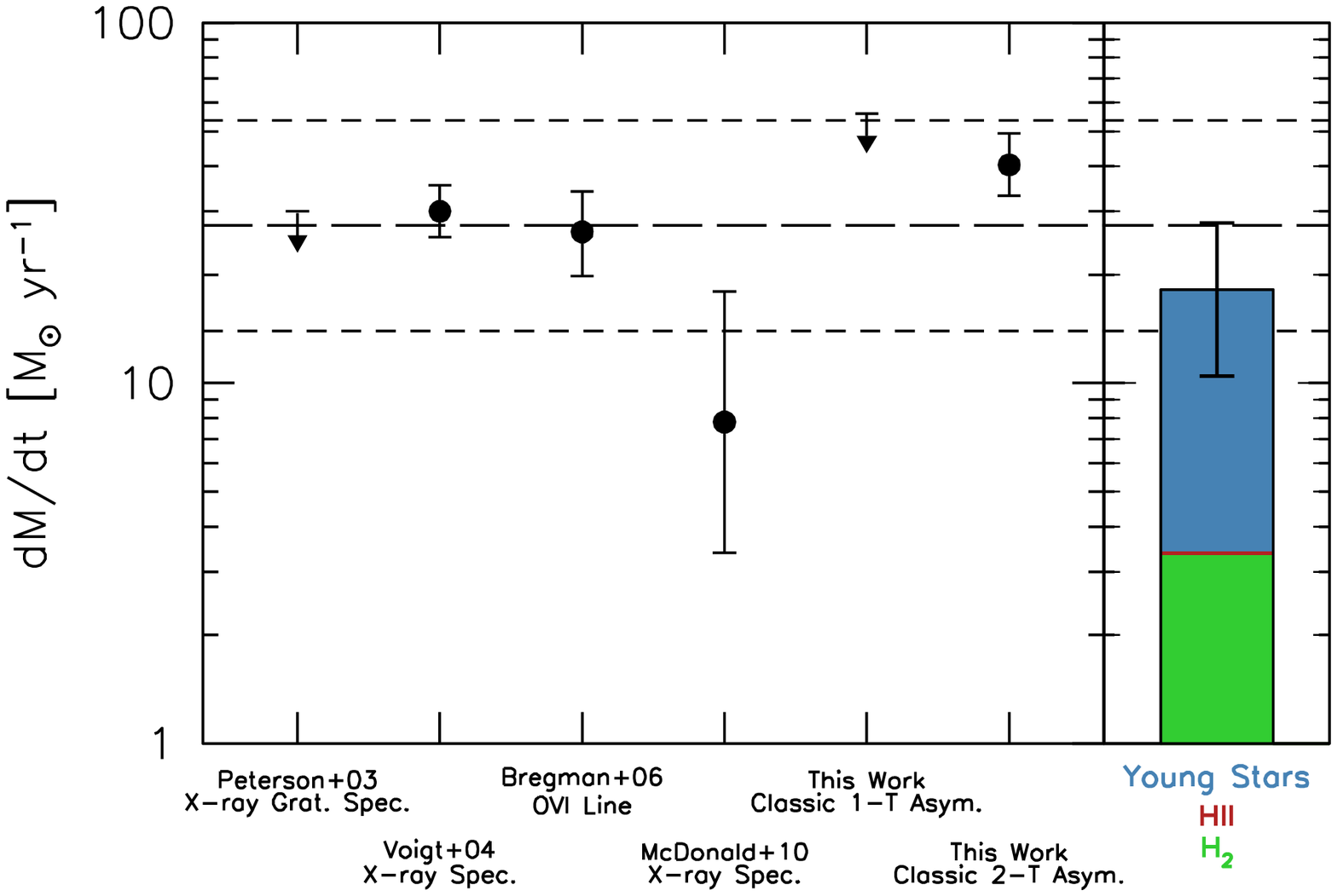}
\caption{ICM cooling rate estimates for Abell1795 from the literature and this work. We use the classical definition of dM/dt, including only the X-ray emission spatially-coincident with the H$\alpha$-emitting filaments. 
The low estimate from \cite{mcdonald10} is due to the exclusion of the central AGN-dominated 2$^{\prime\prime}$ (2.55 kpc). Horizontal dashed lines represent the mean cooling rate of 27$^{+27}_{-13}$ M$_{\odot}$ yr$^{-1}$. The right panel shows the implied cooling rate from the total amount of mass in young stars \citep{mcdonald11b}, cold molecular gas (this work), and warm ionized gas \citep{mcdonald10}, assuming a central (r $<$ 20 kpc) cooling timescale of 0.9 Gyr, Galactic X$_{CO}$, and conservative intrinsic reddening of E(B-V)=0.15. The fractional (contribution from each phase is shown in blue (57\%), green (42\%), and red (1\%). The error bar represents a range of cooling timescales from 0.5--2.5 Gyr combined with uncertainties in each mass estimate. The large amount of molecular gas found at large radii in this system appears to relieve the tension between the cooling estimates and the amount of cooling byproducts.}
\label{dmdt_lit}
\end{figure*}

The second panel of Figure \ref{edge} compares the molecular gas mass to the \emph{classical} cooling rate, (dM/dt)$_{classical}$~$={\frac{2L\mu m_p}{5kT}}$, which is based on the X-ray luminosity. Estimates of dM/dt based on this formula are typically orders of magnitude higher than the observed cooling rate from X-ray spectroscopy \citep[e.g., ][]{peterson03,peterson06}. \cite{edge01} showed that M$_{H_2}$ is correlated with the classically-derived X-ray cooling rate, but that there is much less molecular gas than inferred by reasonable estimates of the cooling time ($\sim$10$^9$~yr). Indeed, when the entire cool core of A1795 is considered, the amount of H$_2$ detected is $<$10\% of the amount which should have cooled in a Gyr. However, if only the strongly-cooling regions are considered (i.e. those which exhibit far-UV and H$\alpha$ emission) this fraction increases to $\sim$30\%. That is, if the observed molecular gas cooled in situ over the past Gyr, it represents $\sim$30\% of the \emph{local} expected classical cooling flow. This large fraction suggests that, when the other cooling byproducts are included, there may not be a significant discrepancy between the classical cooling estimate and the total mass of cooling byproducts \emph{on small scales}.


The apparent balance between the cooling rate and the amount of cooling byproducts is further demonstrated in Figure \ref{dmdt_lit}. We compile estimates of the cooling rate based on X-ray and EUV spectroscopy from the literature \citep{peterson03, voigt04, bregman06, mcdonald10}, and find a mean of 27$^{+27}_{-13}$ M$_{\odot}$ yr$^{-1}$. The 
large uncertainty reflects the variety of methods used. We show, as an upper limit, the estimate of the classical cooling rate in regions with evidence for multiphase (H$\alpha$, CO) gas. As an approximate correction for projection, we model the X-ray spectrum in this region with a two-temperature plasma and calculate the cooling rate from the lower-temperature component. In \cite{mcdonald10}, we calculate the central cooling time (0.9 Gyr) based on a deprojected X-ray spectrum of the central 20~kpc, assuming $t_{cool}$ $\sim$ 10$^8$$\cdot$(K/10 keV cm$^2$)$^{3/2}$$\cdot$(kT/5 keV)$^{-1}$ Gyr, and show that both the central region and the cooling filaments have orders of magnitude shorter cooling times than the surrounding ICM.  Assuming this cooling time of 0.9 Gyr, Figure \ref{dmdt_lit} shows that the total cooling rate of 18.1 M$_{\odot}$ yr$^{-1}$, inferred by young stars \citep[10.4 $\pm$ 2 M$_{\odot}$ yr$^{-1}$;][corrected for extinction]{mcdonald09}, cold molecular gas (7.6 $\pm$ 0.7 M$_{\odot}$ yr$^{-1}$), and warm ionized gas \citep[0.10 $\pm$ 0.05 M$_{\odot}$ yr$^{-1}$;][assuming case B recombination and a large range of electron densities]{mcdonald10} closely matches the estimates of the ICM cooling rate.  These estimates confirm earlier work which showed that the H$\alpha$-emitting gas does not contribute appreciably to the total mass of cooling byproducts in Perseus \citep{johnstone07}. This cooling estimate necessarily hinges on our assumptions of a 0.9 Gyr cooling time and conservative reddening of E(B-V) = 0.15 \citep[50\% of observed Balmer reddening;][]{calzetti94,mcdonald12a}. We show in Figure \ref{dmdt_lit} the uncertainty due to allowing a range of cooling times from 0.5 Gyr (dense filaments) to 2.5 Gyr (diffuse gas in central 60 kpc), and two separate star formation estimates including one with an implicit reddening correction \citep{rosa-gonzalez02} and without \citep{kennicutt98}. 

This good agreement between i) the classical, L$_X$-derived, cooling rate and the spectroscopic cooling rate \emph{in the densest regions of the ICM}, and ii) the spectroscopic cooling estimates and the amount of cooling byproducts, suggests that the cooling flow problem may be much less drastic than previously thought in these dense regions. 

\section{Summary and Conclusions}
We present CARMA observations of the cool core of Abell~1795 at CO(1-0). We find significant amounts of cold molecular gas at clustercentric radii of 0 kpc (3.9 $\pm$ 0.4 $\times$10$^9$ M$_{\odot}$) and 30 kpc  (2.9 $\pm$ 0.4 $\times$10$^9$ M$_{\odot}$), and place an upper limit on the H$_2$ mass at 60 kpc ($<$0.9 $\times$10$^9$ M$_{\odot}$), assuming a Galactic value of X$_{CO}$. These CO(1-0) peaks are coincident in both position and velocity with the warm, ionized filaments and the brightest star-forming regions. The large amount of cold molecular gas at large radius along the cooling filament implies that the ICM is able to cool efficiently far-removed from the effects of the AGN.  In light of these new observations, we compare the ICM cooling estimates to the mass of cooling byproducts and find agreement in the central 10~kpc and along the rapidly-cooling filaments. Furthermore, we find that in these dense regions the local classical cooling rate is in good agreement with both the spectroscopic cooling rate and the cooling rate implied by the cooling byproducts. These results suggest that the cooling flow problem, at least in this system, stems from a lack of observable cooling in the more diffuse (off-filament) regions at radii larger than $\sim$ 10 kpc. In the near future, ALMA observations will allow extended CO emission to be mapped in a large sample of cool cores, shedding further light on the extent of the cooling flow problem.

\section*{Acknowledgements} 
MM was supported by NASA through contract 2834-MIT-SAO-4018, issued by CXO under contract NAS8-03060. SV acknowledges support from NSF through contracts AST-0606932, 100958, and from a Senior NPP Award held at the NASA GSFC. CARMA development and operations are supported by NSF and partner universities. We thank R.~Mushotzky and L.~Mundy for helpful discussions. Finally, we thank the anonymous referee for his/her careful review, which has greatly improved the final manuscript.


\end{document}